# Geomechanics Contribution to $CO_2$ Storage Containment and Trapping Mechanisms in Tight Sandstone Complexes: A Case Study on Mae Moh Basin

Romal Ramadhan[1], Khomchan Promneewat[2], Vorasate Thanasaksukthawee[1],
Teerapat Tosuai[1], Masoud Babaei[3], Seyyed A. Hosseini[4],
Avirut Puttiwongrak[5], Cheowchan Leelasukseree[1], and Suparit Tangparitkul[1,*]

1. Department of Mining and Petroleum Engineering, Faculty of Engineering, Chiang Mai University, Chiang Mai, Thailand
2. Faculty of Civil Engineering Sciences, Graz University of Technology, Graz, Austria
3. Department of Chemical Engineering, The University of Manchester, Manchester, UK
4. Bureau of Economic Geology, Jackson School of Geosciences, The University of Texas at Austin, Austin, TX, USA
5. Geotechnical and Earth Resources Engineering, School of Engineering and Technology, Asian Institute of Technology, Pathum Thani, Thailand

*To whom correspondence should be addressed: suparit.t@cmu.ac.th Tel. +66 53 944 128 Ext 119

**Abstract**

Recognized as a not-an-option approach to mitigate the climate crisis, carbon dioxide capture and storage (CCS) has a potential as much as gigaton of $CO_2$ to sequestrate permanently and securely. Recent attention has been paid to store highly concentrated point-source $CO_2$ into saline formation, of which Thailand considers one onshore case in the north located in Lampang – the Mae Moh coal-fired power plant matched with its own coal mine of Mae Moh Basin. Despite a large basin and short transport routh from the source, target sandstone reservoir buried at deeper than 1000 m is of tight nature and limited data, while question on storing possibility has thereafter risen. The current study is thus aimed to examine the influence of reservoir geomechanics on $CO_2$ storage containment and trapping mechanisms, with co-contributions from geochemistry and reservoir heterogeneity, using reservoir simulator – CMG-GEM. With the injection rate designed for 30-year injection, reservoir pressure build-ups were ~77% of fracture pressure but increased to ~80% when geomechanics excluded. Such pressure responses imply that storage security is associated with the geomechanics. Dominated by viscous force, $CO_2$ plume migrated more laterally while geomechanics clearly contributed to lesser migration due to reservoir rock strength constraint. Reservoir geomechanics contributed to less plume traveling into more constrained spaces while leakage was secured, highlighting a significant and neglected influence of geomechanical factor. Spatiotemporal development of $CO_2$ plume also confirms the geomechanics-dominant storage containment. Reservoir geomechanics as attributed to its respective reservoir fluid pressure controls development of trapping mechanisms, especially into residual and solubility traps. More secured storage containment after the injection was found with higher pressure, while less development into solubility trap was observed with lower pressure. The findings reveal the possibility of $CO_2$ storage in tight sandstone formations, where geomechanics govern greatly the plume migration and the development of trapping mechanisms.

**Keywords**: carbon capture and storage (CCS); $CO_2$ geological storage; reservoir simulation; geomechanics; $CO_2$ storage containment; $CO_2$ trapping mechanisms

**Graphical Abstract**

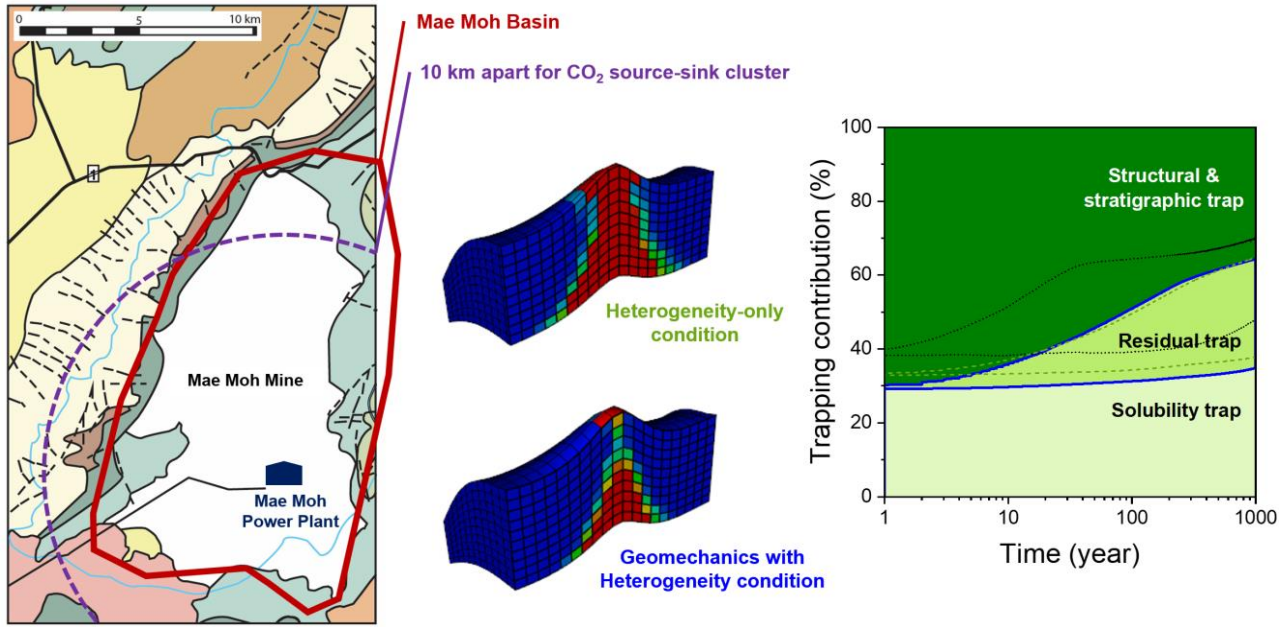

**Highlights**

- A tight sandstone formation was examined for $CO_2$ storage potential.
- Spatiotemporal $CO_2$ plume migration was observed since injection for a millennium.
- Geotechnical influence on storage containment and trapping mechanisms were realized.
- Anticline-shaped and syncline-shaped complexes were compared.
- Integrated influence of geochemistry, heterogeneity, and geomechanics was assessed.

# 1 Introduction

Transitioning away from fossil fuels has been agreed in the recent COP28 in the United Arab Emirates, nevertheless this energy transition is to be delivered through a just, orderly, and equitable manner with goal to achieve the net zero by 2050.[1] The first global stocktake from the COP28 also recognizes actions to accelerate in this critical decade, including, the carbon dioxide ($CO_2$) capture and storage (CCS), particularly in 'hard-to-abate' sectors.[1,2] Such a CCS approach is based on the best available science,[3] which is proven, though scaling-up to a global magnitude at high certainties on storage containment and secured integrity is of challenge.[4]

With limited storage resources in depleted hydrocarbon fields while more challenging affords needed for unconventional alternatives (e.g., salt caverns and coal seams), saline formations scattering around the world are thus anticipated for storing 'gigaton-scale' anthropogenic $CO_2$ released since the industrial revolution.[4,5] Storage performance in saline formations relies on various factors, including (i) fluid-rock geochemical reactions,[6,7] (ii) reservoir heterogeneity,[8,9] and (iii) geomechanical influences,[8,10–12] of which attribute to how the $CO_2$ trapping mechanisms develop.[8,13–16] While previous studies paid much attention to geochemistry-induced trapping mechanisms,[17,18] holistic consideration thereof including geomechanics is usually overlooked and is of interest in the current study.

While geochemical reactions between $CO_2$-dissolved brine and hosting rock could induce both mineral dissolution and precipitation, depending on brine species, minerals, system acidity, and others,[19] leading to either improving storage containment or damaging reservoir and injectivity,[20] reservoir heterogeneity however likely benefits storage containment and security owing to $CO_2$ flow hindrance in vertical direction and fluctuated local capillarity.[21] Considering tight sandstone reservoirs, which are widely abundant with prime properties of strong acid resistance and fair heterogeneity, geomechanics contribution to the trapping mechanisms in association with $CO_2$ plume migration and reservoir fracture resistance could be thus a crucial factor due to nature strength of consolidated sandstone.[22,23] Previous studies on reservoir heterogeneity, relevant geochemistry, and reservoir geomechanics were researched with resulted $CO_2$ storage behaviors, which are concluded in **Table 1**.

Table 1. Previous studies on reservoir contributing factors and the resulted $CO_2$ storage behaviors.

| Reservoir contributing factor | Resulted $CO_2$ storage behavior |
|---|---|
| **Heterogeneity** (porosity and permeability) | $CO_2$ plume migration<br>• Fang et al. found that when permeability discrepancy is relatively large, $CO_2$ plume preferentially migrates along horizontal layer without vertical direction.[24]<br>• Al-Khdheeawi et al. found that homogeneous reservoir has a greater vertical migration distance of $CO_2$ plume when compared to heterogeneous one.[25]<br>Trapping mechanisms<br>• Rasheed et al. examined how various Lorenz coefficients may affect the degree |

| Reservoir contributing factor | Resulted $CO_2$ storage behavior |
|---|---|
| | of heterogeneity to better understand the impact of heterogeneity on trapping mechanisms and unveiled that a low-to-medium-level heterogeneous reservoir (with adequate porosity) may be a promising option for $CO_2$ storage as it increases solubility trapping.[26]<br>• Gershenzon et al. conducted research on the impact of small-scale heterogeneity on $CO_2$ trapping processes in deep saline aquifers. They discovered that variations in capillary pressure entry points for various materials can cause $CO_2$ to be trapped in heterogeneous media. They came to the conclusion that capillary trapping mechanisms in a highly heterogeneous reservoir may significantly outperform those in a less heterogeneous reservoir.[27] |
| | Final Pressure<br>• Rasheed et al. also found an impact of heterogeneity on reservoir pressure as the reservoir pressure at the end of injection decreases with increasing heterogeneity.[26] |
| **Heterogeneity** (capillary pressure) | $CO_2$ plume migration<br>• Jackson and Krevor performed research on heterogeneity in small-scale capillaries associated with plume migration and discovered that lateral migration rates may be increased by 200% in layered heterogeneities.[28] |
| | Trapping mechanisms<br>• Harris et al. examined an effect of heterogeneity in capillary pressure on capillary or residual trapping, and found that an increase in heterogeneity leads to an increase in capillary trapping by three times larger.[22] |
| **Heterogeneity** (wettability) | $CO_2$ plume migration<br>• Al-Khdheeawi et al. found that wettability heterogeneity significantly increases vertical $CO_2$ plume movement, which has a large impact on dissolution and residual trappings.[29] |
| | Trapping mechanisms<br>• Al-Khdheeawi et al. also found wettability variability influencing on trapping capacity, and discovered that heterogeneous wettability decreased residual trapping but enhanced solubility trapping, increasing a quantity of transportable $CO_2$.[29] |
| **Geochemistry** | Trapping mechanisms<br>• Nghiem et al. examined the impact of geochemistry on mineral entrapment and discovered that calcite, siderite, and dolomite are created in around 1 – 2 moles after 100 moles of injected $CO_2$.[10] |
| | Storage capacity<br>• Chidambaram et al. discovered a 6% reduction in storage capacity when considering geochemistry effect.[30] |
| **Geomechanics** | Formation uplift<br>• Khan et al. discovered that, in the absence of a reservoir fault, ground uplift will peak just above the $CO_2$ injection port, but in the presence of a geological fault, the ground uplift will peak just above the $CO_2$ leakage point.[31]<br>• Jun et al. examined the uplift of Pohang Basin and Donghae gas reservoir, and discovered maximum uplift of 25.4 and 32.6 mm, respectively, using the Gaussian pressure transient approach.[32] |
| | Fault reactivation<br>• In Snøhvit project, Chiaramonte conducted a study on geomechanics and discovered that the critical pressure perturbation required for reactivation is over 13 MPa, the limiting pressure rise before to reaching the fracture pressure.[33]<br>• Vilarrasa et al. investigated how the site of $CO_2$ injection affected fault stability, the authors discovered that $CO_2$ injection reduced fault stability. Therefore, to |

| **Reservoir contributing factor** | **Resulted CO$_2$ storage behavior** |
|---|---|
| | reduce fault stability concerns and prevent imposing injection rate restrictions, injection wells should be placed as far away from faults as feasible.[34] |
| | Induced seismicity <br> • In Oklahoma in 2011 and 2012, Song et al. summarized one of the most well-known instances of produced seismicity using fluid injection into the earth. The injection procedure at Oklahoma's disposal wells caused pore pressure changes, which resulted in seismic activity with a magnitude of 5.6.[35] |
| | Reservoir stability <br> • Khan et al. investigated how reservoir stability in the Biyadh and Minjur reservoirs is impacted by geomechanics. For the purpose of preventing fault activation and caprock collapse, the maximum injection pressures were found to be 27 and 56 MPa, respectively.[31,36] |

Despite the fact that the CO$_2$ trapping mechanisms are well-defined and increasingly investigated in recent years, simultaneous influences of such intricate interplay among the three factors are still infancy and need further exploration.[34,37] In consecutive order of their temporal developments and contribution to storage security, the trapping mechanisms consist of (i) structural and stratigraphic, (ii) residual, (iii) solubility, and (iv) mineral traps.[38,39] With further understanding on such a geochemistry-heterogeneity-geomechanics 'trio', the storage containment as attributed from the trapping mechanisms could be emphasized and even engineered to assure a 'permanent' and 'secure' storage.[23,30] The current study therefore endeavors to thoroughly examine such a geochemistry-heterogeneity-geomechanics 'trio' (of which previous studies often neglected a contribution from geomechanics), with anticipation to shed some lights on how this triple-contribution is crucial on the CO$_2$ storage in the tight sandstone. Novel contribution is to offer insights into viability, challenges, and optimization strategies that are associated with such reservoir condition, of which might be important for scaling up the global storage quantity to achieve gigaton scale.

In the current study, an interplay among the trio is examined on a geological setting of tight sandstone complexes, based on a prospect candidate of on-shore storage site in northern Thailand – the Mae Moh Basin in Lampang.[40] Two representatives of reservoir complexes are considered, namely anticline and syncline structures, to thoroughly assess all possible geological structures of the basin. With a well-constructed reservoir simulator, the current work aims to elucidate on how the trio contribute to the storage trapping mechanisms and hence the storage containment holistically, where insightful emphases on storage integrity are to be highlighted. In support of CCS implementation in the area and furthermore in similar kind of storing reservoir conditions, possibility or potentiality of the basin on such studied scopes are to be also addressed.

## 2 Geological Setting and Simulation Methodology

### 2.1 Field background and storage location

To comply with Thailand's carbon neutrality target in 2050, recent policy on electricity generation in Thailand has set to be carbon neutral by 2050 accordingly by the Electricity Generating Authority of Thailand (EGAT).[41] One of EGAT power plants providing an electricity to northern Thailand is a coal-fired power plant located in Mae Moh district of Lampang, where substantial lignite deposits (i.e., Mae Moh mine) are nearby and have been mined since 1954.[42] Due to its close distance (~10 km) between the power plant and its own mine, possibility to store $CO_2$ from the power plant to the mine (of Mae Moh Basin) is of interest, see **Fig. 1**, with aim to store as much as 15 Mtpa $CO_2$ released from the power plant until the coal reserves ceased (~300 Mt $CO_2$ in ~30-year period).

Despite being mined for more than 60 years, only surface geology of Mae Moh Basin where lignite deposited (~300 m) is well characterized, whereas deeper underground formations were left unattended. EGAT's own geology division has recently deduced geological cross-section of the basin deep formations, which is employed for generating a block model of numerical analysis in the current study (will be discussed in the following part). According to Chaodumrong,[43] the rock formations from the deeper order in Mae Moh Basin are Permo-Triassic volcanic rock (PmTr), coarse- to fine-grained red siliciclastics (Tr1), massive limestone (Tr2), fine-grained turbidites with grey mudstone (Tr3), and reddish gray argillaceous limestone (Tr4). Owing to general $CO_2$ storage prerequisites,[44,45] Tr1 as a promising sandstone formation buried more than 800 m underneath is chosen as a target reservoir for storage, covered with Tr2 acting as seal or cap rock in the current study.

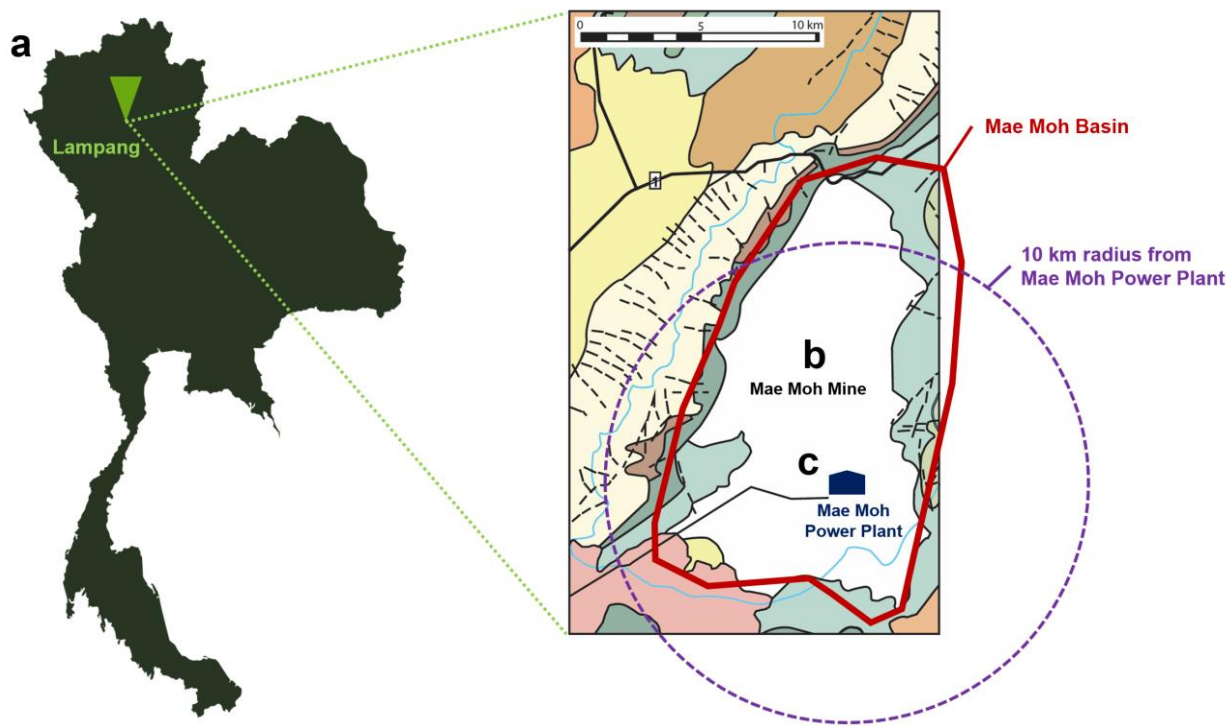

**Figure 1.** Mae Moh mine (b) sitting on Mae Moh Basin in northern Thailand (a) is located ~10 km from its coal-fired power plant (c) where $CO_2$ released ~15 Mtpa. Figure adapted from Thanasaksukthawee et al.[40]

**2.2 Numerical model description**

A series of numerical simulations were conducted using a commercial compositional reservoir simulator developed by Computer Modelling Group Ltd. (Canada), i.e., CMG-GEM.[46,47] CMG-GEM simulates fluid flow and relevant phenomena in reservoir under a spatiotemporal discretization of material and energy balance equations through finite volume and finite difference methods.[48] To simulate the $CO_2$ trapping mechanisms, the CMG-GEM employed a generalized Peng-Robinson equation of state to forecast and evaluate phase equilibrium, solubility, and thermodynamic characteristics of $CO_2$ within geological formations throughout an injection stage.[49] Governing equations used in the simulation are reported in **Section S1** in **Supplementary Material**.

Two types of reservoir structural complexes are designed: (i) anticline-shaped and (ii) syncline-shaped structures, principally based on Tr1 formation to facilitate an intricate understanding of cross-sectional dynamics within the basin. A sequential Gaussian with normal distribution was selected to assess the uncertainty inherent in Tr1 reservoir, with porosity ($\phi$) and log-permeability ($\log k$) geo-statistically defined and the two are linearly correlated.[50] Reconstructed reservoir complexes of a 1 km² are shown in **Fig. 2**, with porosity (**Figs. 2a** and **2c**) and permeability (**Figs. 2b** and **2d**) distributions annotated. The $\phi$ and $k$ ranges are from outcrop tests (**Section S2** in **Supplementary Material**), taken to be 1.6 – 2.3% and 0.27 – 8 mD, respectively. Vertical to horizontal permeability ratio is taken to be 0.1.[51] Structural dip angle of 10° was assumed for reservoir reconstruction.[52]

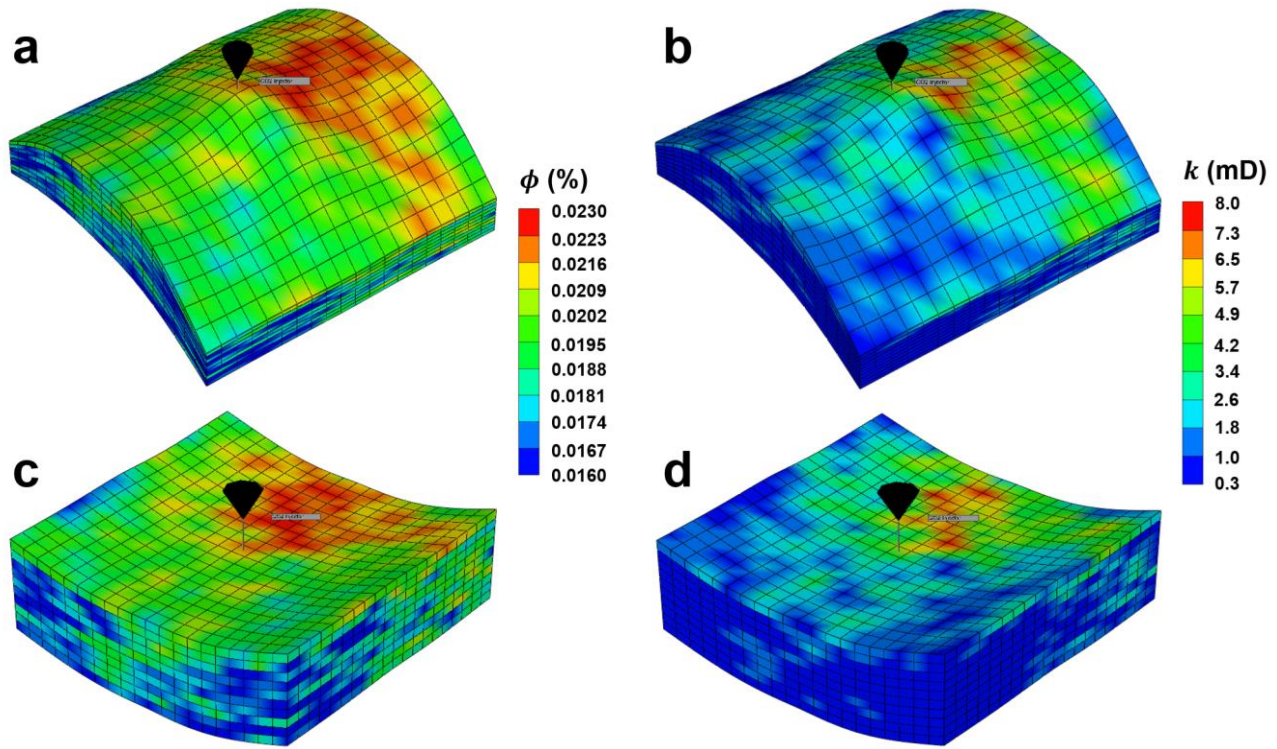

**Figure 2.** Reservoir structural complexes: anticline-shaped (a and b) and syncline-shaped (c and d), with geo-statistically distributed porosity ($\phi$: a and c) and permeability ($k$: b and d) for heterogeneous condition, of tight sandstone models used in the current study. $CO_2$ injection is positioned at the center of the reservoir, with inverted cone symbols annotated.

Reservoir properties and characteristics of the two complexes are summarized in **Table 2**, of which the top part of the upper anticline complex is set at 1020 m depth and of the lower syncline complex is at 2300 m, according to the geology reference. Reservoir pressure and temperature gradients were assumed based on Thailand's averages of 9.523 kPa/m and 30 °C/km, respectively.[53,54] Reservoir pressure and temperature used for simulation were estimated at the top of each reservoir complex (i.e., minimal pressure), ensuring storage containment and security.

**Table 2.** Reservoir properties and characteristics used in the simulation study.

| Properties and characteristics | Reservoir complex | |
|---|---|---|
| | Anticline-shaped | Syncline-shaped |
| Top depth (m) | 1020 | 2300 |
| Grid dimensions | 20 × 20 × 10 | 20 × 20 × 10 |
| Reservoir size (m) | 1000 × 1000 | 1000 × 1000 |
| Reservoir thickness (m) | 370 | 370 |
| Reservoir pressure (MPa) at the top depth | 9.30 | 22.93 |
| Reservoir temperature (°C) at the top depth | 57 | 88 |

To thoroughly observe $CO_2$ plume migration and associated storage containment, the injection position is located at the center of reservoir and perforated at the lowest depths of each complex (1353-1390 m and 2633-2670 m for anticline-shaped and syncline-shaped, respectively). Perforation height of ~22.6 m is made at the lowest grid of each complex.

Three conditions of reservoir simulations were considered, with geochemistry influence included in all cases and saline composition reported in the next section. Homogeneous condition refers to invariant reservoir, with $\phi$ of 1.95% and $k$ of 1.357 mD taken from the parameter distribution averages discussed above (**Figs. S1 – S3**), of which define the heterogeneous simulating condition. Heterogeneous with geomechanics condition assumes both geo-statistical $\phi$-$k$ variation and reservoir geomechanics as described in the following section. Due to limited field-scale data of Mae Moh Basin, reservoir boundary is assumed to be a closed or no-flow system.

3D simulation models were conducted to observe all trapping mechanisms and their contributions to $CO_2$ storage security in a 1000-year time frame. Spatiotemporal patterns of $CO_2$ plume migration were also monitored. $CO_2$ was continuously injected into saline reservoir (Tr1) for 30 years, complying with the anticipated period of the CCS project at Mae Moh area.[41] For 30 years, the anticline complex is set to receive a constant injection at 105 tons per day, while the syncline is set to experience a higher injection rate of 1350 tons per day. The two different injection rates are designed to achieve maximum storage capacity without exceeding their respective fracture pressure (discussed in **Section 3.1**), validated by sensitivity analysis on various injection rates (**Fig. S4**). **Figure 3** illustrates the numerical simulations studied in the current work, including main $CO_2$ storage and reservoir behaviors and sensitivity analysis.

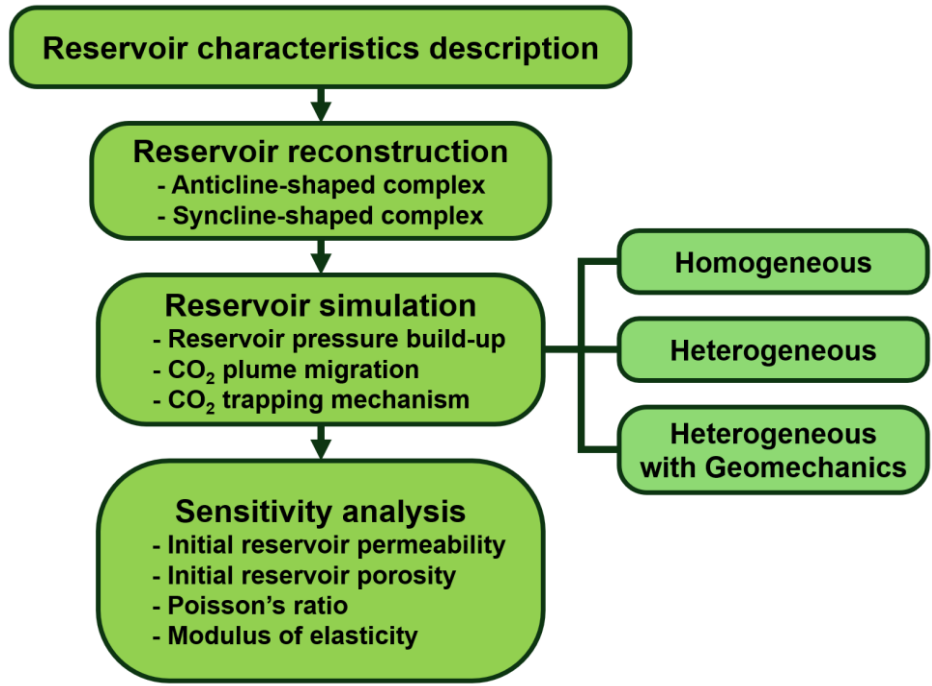

**Figure 3.** Flowchart summarizes numerical simulations performed in the current study, including main $CO_2$ storage and reservoir behaviors and sensitivity analysis.

**2.3 Geomechanical properties and saline composition**

Geomechanical properties of storage complexes used for simulation in the current study were taken from the report by Chaodumrong,[45] including those of reservoir (Tr1) and seal (Tr2) rocks, and are reported in **Table 3**. In the simulation, initial minimum horizontal stress ($\sigma_h$) and initial vertical stress ($\sigma_V$) were estimated by using $\sigma_h = 0.0184Z$, where $Z$ is the vertical depth, and taking the overburden gradient of 23.1 MPa/km from Zhang,[55] which resulted in the $\sigma_h$ of 18.8 MPa and 42.3 MPa, and the $\sigma_V$ of 23.5 MPa and 53.1 MPa, for anticline-shaped and syncline-shaped complexes, respectively. Such results agree with a study by Tingay et al.,[56] who found a relation of the three principles stresses in Thailand as $\sigma_H \approx \sigma_V > \sigma_h$.

Table 3. Geomechanical properties of storage complexes used in the current study.

| Geomechanical properties | Formation | |
| --- | --- | --- |
|  | Tr1 (Tight sandstone reservoir) | Tr2 (Limestone seal) |
| Compressive strength (MPa) | 75 | 75 |
| Young modulus (GPa) | 10.7 | 35.1 |
| Poisson's ratio | 0.25 | 0.22 |
| Cohesion (MPa) | 6.31 | 4.49 |

A reservoir stability analysis was not performed in the current work since the reservoir pressures were designed to be not exceed ~80% of the known fracture pressures for both reservoir complexes (**Fig. S4**), a precautionary measure aimed for a safety margin and mitigating the risk of reservoir instability. The reservoir characters studied are also of inherent strength and coupled with the limited pressures applied, securing sufficient stability. Previous study in low-permeable (40 mD) formation as the current one also confirms no-leakage across storage complex.[34]

Saline composition used for simulation is defined as per sample examination. With total dissolved solids of 1561 mg/L, the dominant salt ions are 472 mg/L sodium ($Na^+$), 16.1 mg/L potassium ($K^+$), 0.993 mg/L calcium ($Ca^{2+}$), 19.1 mg/L magnesium ($Mg^{2+}$), 0.025 mg/L iron ($Fe^{2+}$), 2.73 mg/L fluoride ($F^-$), 4.81 mg/L chloride ($Cl^-$), and 744 mg/L sulfate ($SO_4^{2-}$). Some other ions may present at undetectable amounts, e.g., aluminum and nitrate.

## 3 Results and Discussion

### 3.1 Reservoir pressure build-up and fracture pressure

To examine geomechanical aspect on $CO_2$ storage containment integrity, reservoir pressure build-up due to $CO_2$ injection is simulated and compared against reservoir fracture resistance.[57] Considering the current storages in saline formation, reservoir fracture pressures were estimated based on Hubbert and Willis method[58] with assuming pressure gradient as discussed in **Section 2.2**. Resulted fracture resistances are 17.5 MPa and 35.2 MPa for anticline and syncline complexes, respectively.

When geochemistry is considered, influences of reservoir heterogeneity and geomechanics on reservoir pressure build-up are pronounced and the resulted simulations are shown in **Fig. 4**. In the 30-year injection period, reservoir pressures increased rapidly in all cases considered, with stronger pressure build-up observed when no-geomechanics considered (green-dashed and black-dotted lines in **Fig. 4**). This was due to an increase in cumulative injected $CO_2$ in confined reservoir pore spaces where connate brine initially resided.[59] lower reservoir pressure when considered geomechanics was simply due to Terzaghi's principle,[60] $p' = p_t - p_p$, where $p'$, $p_t$, and $p_p$ are the effective stress, the total stress of reservoir matrix, and the reservoir pore pressure, respectively. In the post-injection period, reservoir pressures were faded gradually and likely stabilized from ~600 years after the injection time, reflecting development toward residual trapping where gravitational force becomes to dominate and brine phase starts to displace 'non-residual' $CO_2$ upward.[39,61] Gradual decrease in reservoir pressure was also contributed to developing in a process of $CO_2$ dissolution,[62] hence a solubility trap.

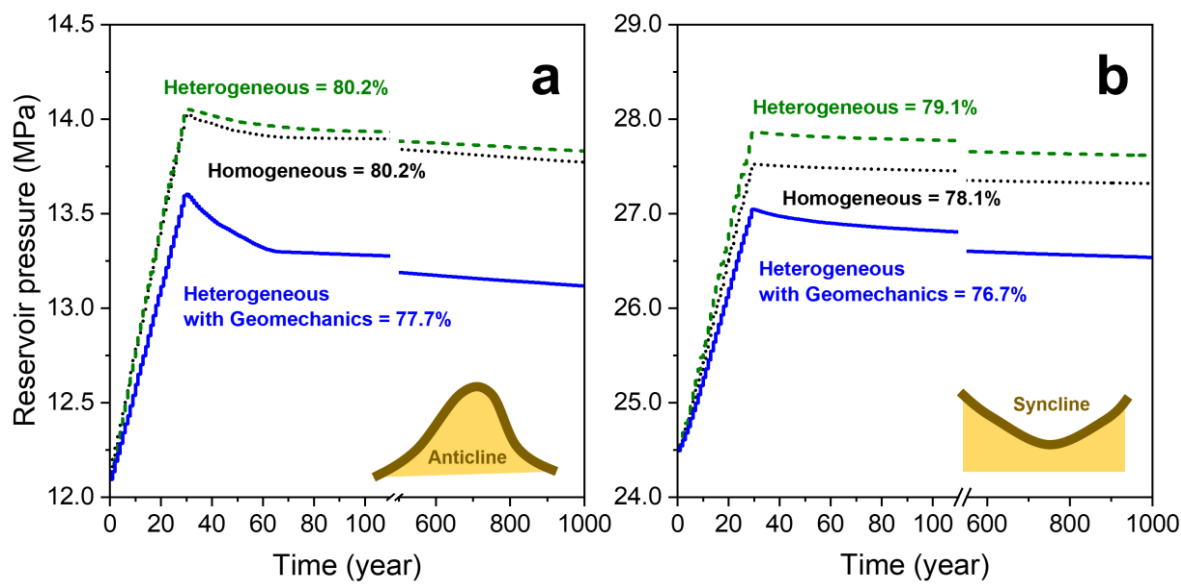

**Figure 4.** Reservoir pressures build-up after continuous injection of $CO_2$ for 30 years, followed by pressure stabilization observed for a millennium after the injection. Reservoir heterogeneity and geomechanics effects are compared: green dashed lines are of heterogenous; black dotted lines are of homogeneous; and blue solid lines are of heterogenous with geomechanics. Percentages indicate the highest reservoir pressures obtained compared to their respective reservoir resistances. Two reservoir complexes are considered: anticline-shaped (a) and syncline-shaped (b).

In the whole studied period for both reservoir complexes, pressure build-up in heterogenous reservoirs was slightly higher than those of homogeneous model. This is due to variations in permeability and porosity that induce limited pathways for $CO_2$ flow, and hence leads to localized accumulation of fluid with higher pressure, whereas in more uniform reservoir the $CO_2$ flows and distributes more evenly with less localized fluid accumulation. Recent study even found a correlation between heterogeneity and reservoir pressure build-up.[63] When the effect of reservoir geomechanics was considered in simulation, the reservoir pressure build-up responses were relatively lower than those of without geomechanics as per discussed above on Terzaghi's principle. Previous study by Kim et al.[64] also confirmed such a result. The authors found that reservoir pore pressure experienced a slight decrease due to overburden included.

With reservoir pressure build-up responses in all conditions being lower than the reservoir fracture resistances estimated (<80% of the fracture resistance at the highest build-up, see **Fig. 4**) in both reservoir complexes, the injection rates designed for the simulation ensure storage security. When considering the 'trio' effects (including geomechanics; blue solid lines in **Fig. 4**), the reservoir pressures due to $CO_2$ cumulative storage at the peak of 30-year injection period are even lower (<77%) as discussed above. This is in line with recommendation for geomechanical stability, where reservoir pressure should not exceed ~75% of fracturing resistance.[65]

## 3.2 CO$_2$ plume migration

Spatiotemporal developments of CO$_2$ plume migration at different conditions are illustrated as CO$_2$ saturation ($S$) on a 1 km$^2$ reservoir complex, shown in **Figs. 5 – 7** and **Figs. 8 – 9** for anticline-shaped and syncline-shaped, respectively.

For an anticline-shaped complex, no CO$_2$ plume reached the top of reservoir in all simulated conditions after a 30-year injection period (**Figs. 7a**, **7e**, and **7i**), reflecting tight sandstone characteristics of low porosity and low permeability. Reservoir heterogeneity appears to benefit CO$_2$ storage containment, with less vertical migration path (~259 m) from the injection point compared to that of homogeneous one (~296 m) where buoyancy-dominant is more pronounced.[66] CO$_2$ plume however appears to migrate laterally at wider distance when heterogeneity is considered (~350 m compared to ~300 m), owing to a vertical to horizontal permeability ratio per se.[21,67] This implies a positive contribution of heterogeneity to storage capacity as found in some previous studies,[68,69] and potentially the consequent residual and solubility traps since CO$_2$-brine contact area is likely increased. When geomechanics influence was included, CO$_2$ plume migrated to a lesser extent in both vertical and horizontal directions due to a strong hindrance of reservoir rock strength onto such a fluid flow (**Figs. 6i** and **7i**).

After the injection had stopped, CO$_2$ plume eventually reached the top of reservoir complex and developed to expand horizontally, see **Fig. 5**. Even though heterogeneity did help to prevent fluid flow upward when compared to homogeneous reservoir (**Figs. 7d** and **7h** at 1000 years after injection), more dramatic hindrance to fluid flow was much obvious with geomechanics contribution – only few CO$_2$ plume traveled to the reservoir top while majority of CO$_2$ was securely confined within the reservoir, see **Figs. 7j – 7l**. CO$_2$ plume migration at reservoir bed was also lesser in expansion when compared to other cases without geomechanics effect, see **Fig. 6**. When geomechanics coupled with geochemistry as realized in the current study, reservoir $\phi$ and $k$ would develop with fluctuation due to mineral dissolution and precipitation over time as observed by Yong et al.[70] This indicates a crucial contribution of geomechanics to CO$_2$ migration behavior and storage containment, especially in such a tight reservoir where geomechanics likely dominates in the current study, of which cannot be neglected when modeling CO$_2$ injection and its associated plume behavior.

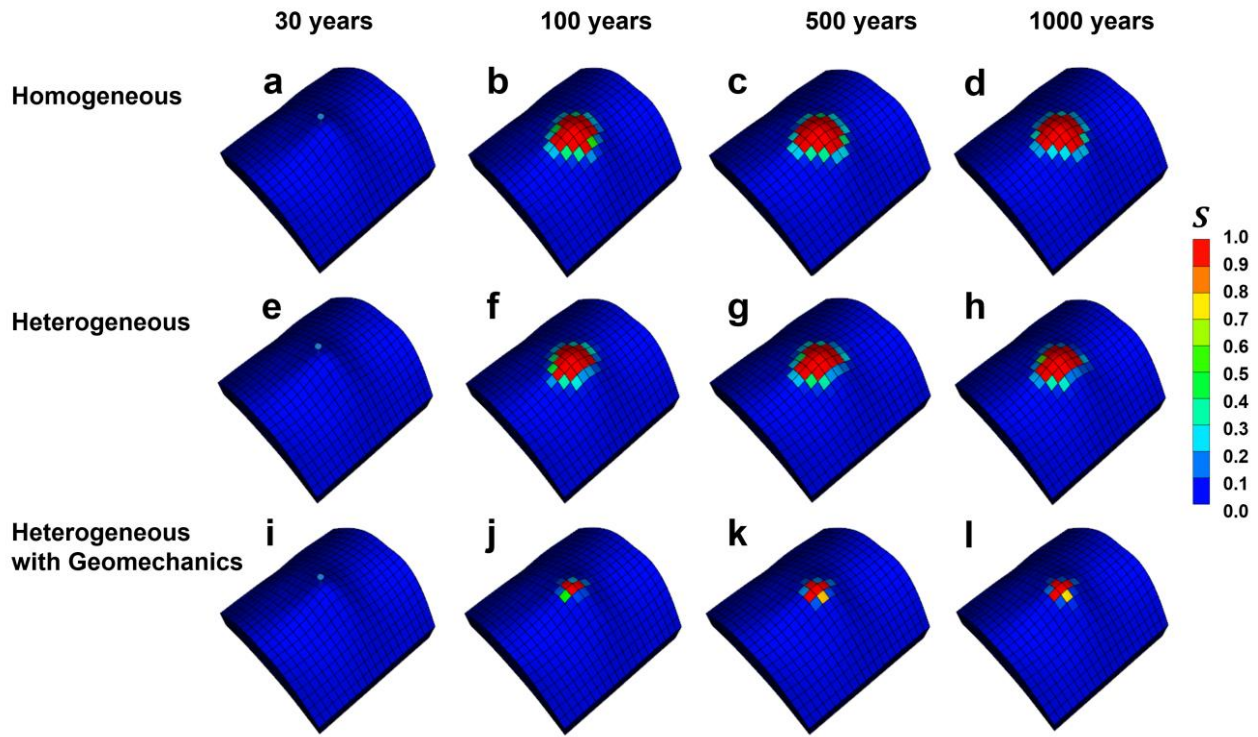

**Figure 5.** Isometric top-views of anticline-shaped reservoir formation: homogeneous (a – d); heterogeneous (e – h); and heterogeneous with geomechanics (i – l), showing $CO_2$ plume migration development over time at 30, 100, 500, and 1000 years, respectively. $CO_2$ saturation ($S$) is indicated by color of the grids.

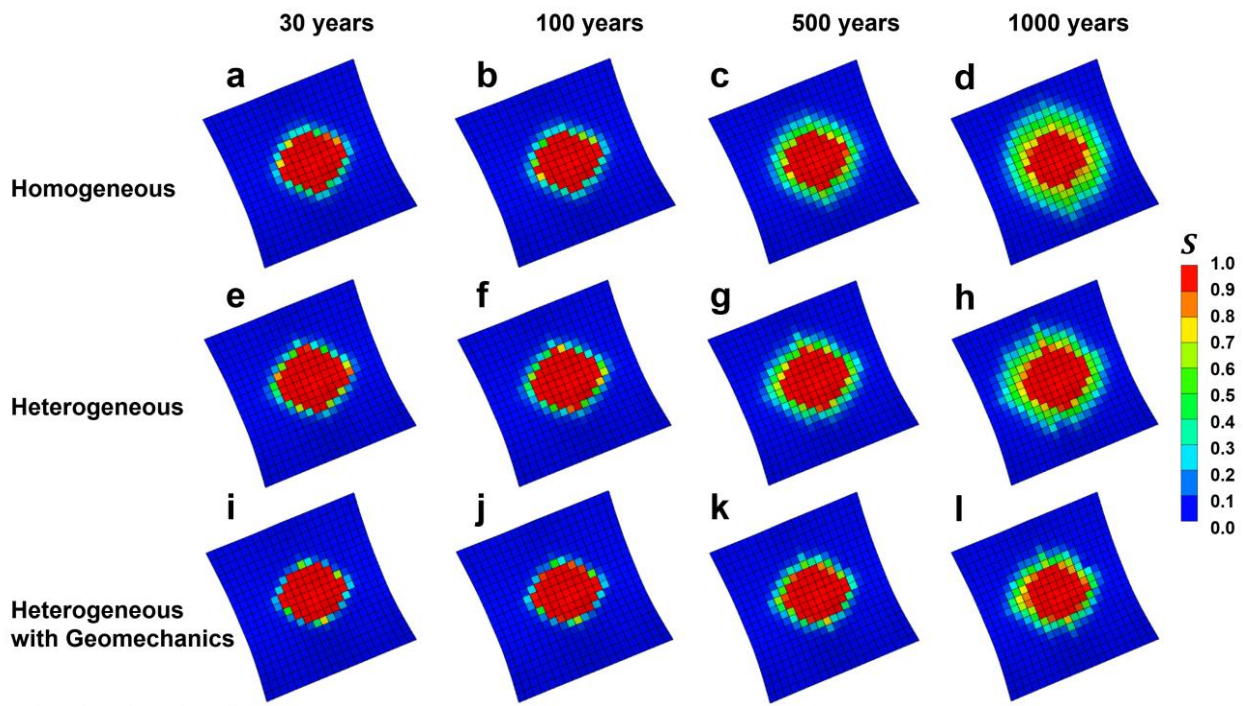

**Figure 6.** Bottom views of anticline-shaped reservoir formation: homogeneous (a – d); heterogeneous (e – h); and heterogeneous with geomechanics (i – l), showing $CO_2$ plume migration development over time at 30, 100, 500, and 1000 years, respectively. $CO_2$ saturation ($S$) is indicated by color of the grids.

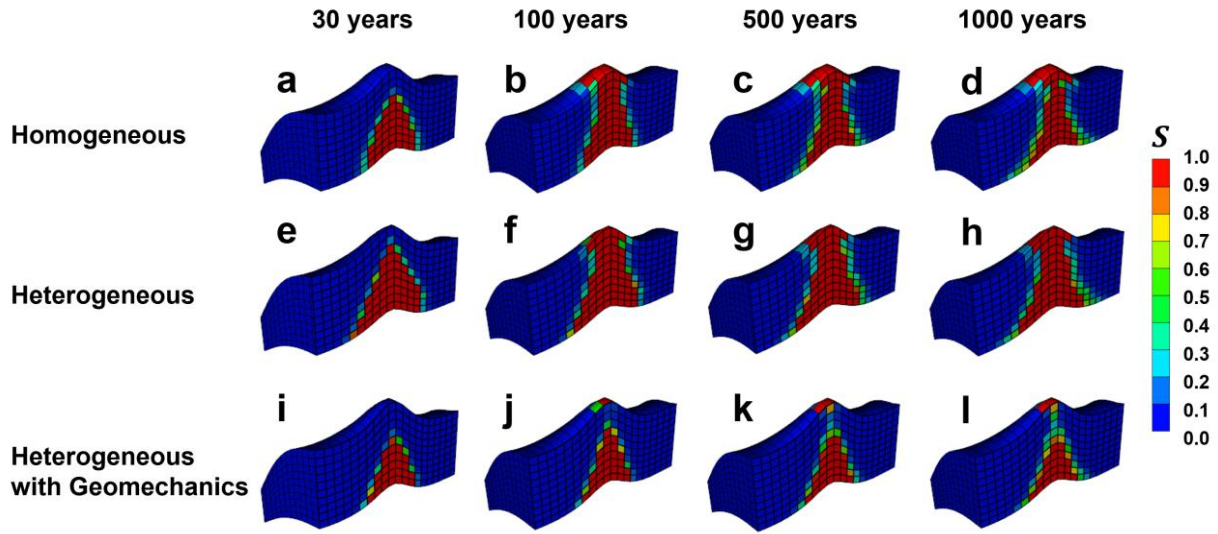

**Figure 7.** Isometric half-width profile-sections of anticline-shaped reservoir formation: homogeneous (a – d); heterogeneous (e – h); and heterogeneous with geomechanics (i – l), showing $CO_2$ plume migration development over time at 30, 100, 500, and 1000 years, respectively. $CO_2$ saturation ($S$) is indicated by color of the grids.

For a syncline-shaped complex, similar behaviors of $CO_2$ plume migration to the anticline-shaped complex were observed (**Figs. 8 – 9**), but no $CO_2$ has reached the reservoir top over a millennium period of simulation (**Figs. 9d**, **9h**, and **9l**). While heterogeneity resisted the fluid flow in both directions, geomechanics-coupled condition resulted in much confined $CO_2$ plume after a millennium (**Figs. 8l** and **9l**). Due to higher reservoir pressure at deeper location, the $CO_2$ plume was predictably smaller in size when compared to the anticline-shaped complex.

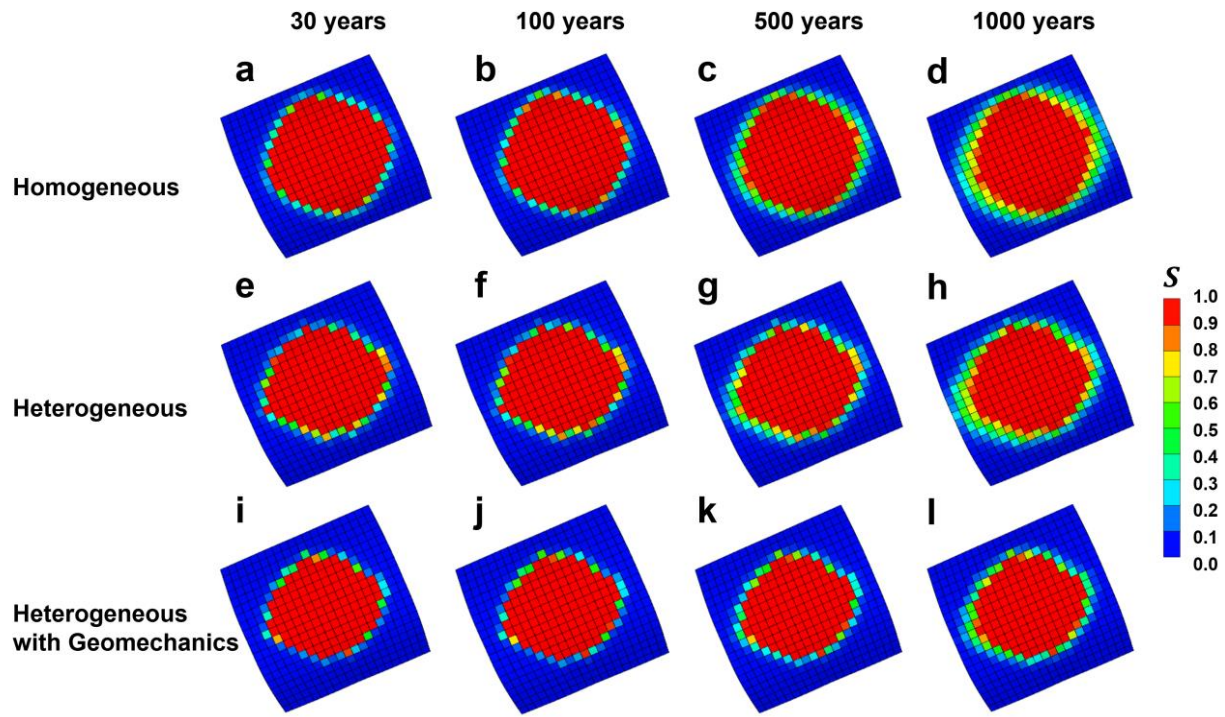

**Figure 8.** Bottom views of syncline-shaped reservoir formation: homogeneous (a – d); heterogeneous (e – h); and heterogeneous with geomechanics (i – l), showing $CO_2$ plume migration development over time at 30, 100, 500, and 1000 years, respectively. $CO_2$ saturation ($S$) is indicated by color of the grids.

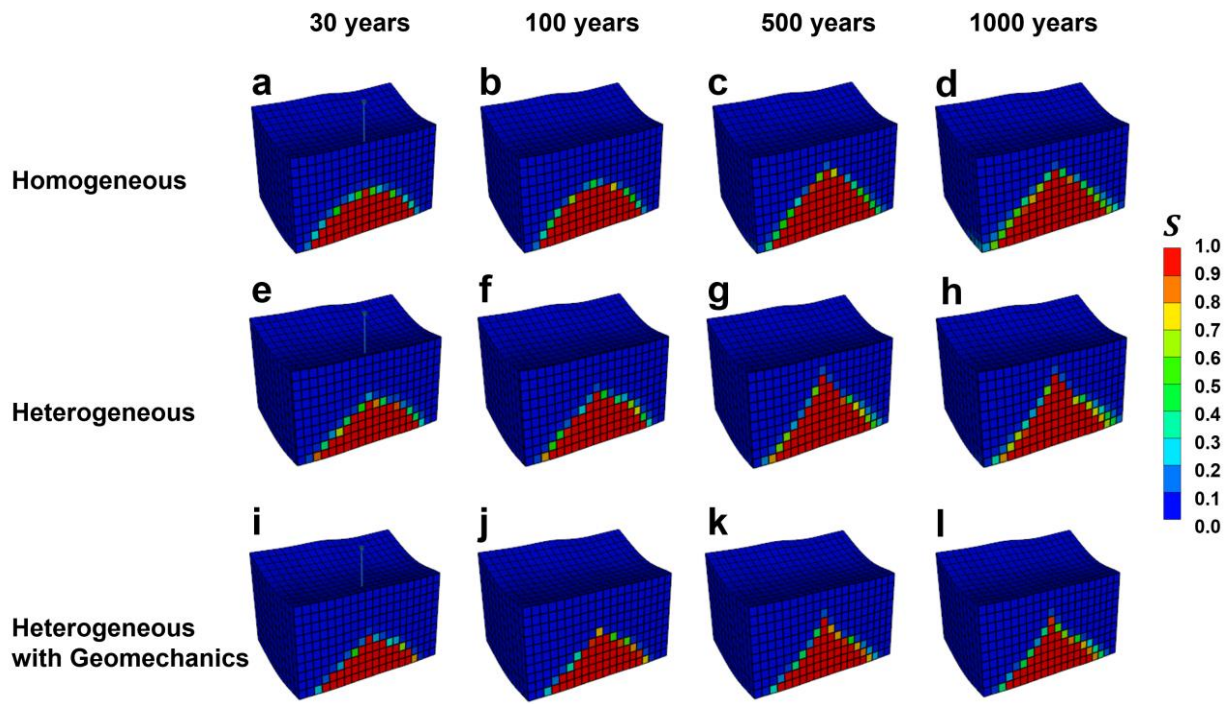

**Figure 9.** Isometric half-width profile-sections of syncline-shaped reservoir formation: homogeneous (a – d); heterogeneous (e – h); and heterogeneous with geomechanics (i – l), showing $CO_2$ plume migration development over time at 30, 100, 500, and 1000 years, respectively. $CO_2$ saturation ($S$) is indicated by color of the grids.

Although precise shape and migration behavior of CO₂ plume at the steady state (i.e., 1000 years) depend on many factors, two main factors principally dominate – the fluid density difference ($\Delta\rho = \rho_{brine} - \rho_{CO_2}$) and the injection rate ($Q$). With larger $\Delta\rho$, the plume likely migrates vertically and spreads at wider distance. On the contrary, the higher $Q$ likely promotes viscous flow and leads to wider region at near wellbore (i.e., the 'dry-out' zone).[71] Assuming a horizontal saline formation, the gravity to viscous ratio ($\Gamma$) and respective maximum plume migrating radius ($r_{max}$) can be determined by:[72]

$$\Gamma = \frac{2\pi\Delta\rho k \lambda_{brine} B^2}{Q} \tag{1}$$

$$r_{max} = \sqrt{\frac{\lambda_{CO_2}}{\lambda_{brine}} \frac{Qt}{\pi B \phi}} \tag{2}$$

where $\lambda_{brine}$ is the brine mobility (ratio of relative permeability to viscosity), $\lambda_{CO_2}$ the CO₂ mobility, $B$ the reservoir thickness, and $t$ the injection period.

For the storage complexes in the current study, the $\Gamma$ are estimated to be $3.2 \times 10^{-14}$ and $1.9 \times 10^{-14}$ for anticline-shaped and syncline-shaped, respectively, with parameters summarized in **Section S4**. Both resulted $\Gamma$ are much lower than 1, indicating a strong viscous-dominated flow. Higher $\Gamma$ in the anticline-shaped complex reflects the greater density difference at lower reservoir depth, compared to the syncline-shaped, see **Table 2** and **Table S1**. This also agrees with the estimated $r_{max}$, where the syncline-shaped has smaller migrating radius (113 m) than its counterpart (177 m) due to less buoyancy at play. Simulated results (**Figs. 7l** and **9l**) are well justified by such theoretical analyses of $\Gamma$ and $r_{max}$ – CO₂ plume appeared to migrate laterally at reservoir bottom rather than float upward. With high $Q$, much pronounced contribution can even be observed at the syncline-shaped complex (**Fig. 9l**).

### 3.3 CO₂ trapping mechanisms and their dynamic contributions

With the simulated results of CO₂ trapping mechanisms obtained (**Fig. 10**), the contributions from each considered factor (namely, heterogeneity and geomechanics) to the trapping mechanisms since the injection stopped can be elucidated and compared. The trapping contribution results are in accordance with the concept reported by Metz et al.,[73] of which the structural and stratigraphic trap contributes at higher degree in the early period, followed by residual and solubility traps come into play at later stage. It is noted that a mineral trap was found negligible in the current studied conditions, though geochemistry has been included.

In the anticline-shaped complex (**Fig. 10a**), a slight difference is observed among the three conditions. In the initial period (<40 years), the contribution of structural and stratigraphic trap changed to be that of residual trap, while that of solubility trap remained the same relatively. After such an initial period, contributions of the two consequent traps increased as expected. Interestingly, when considering geomechanics (blue solid lines in **Fig. 10a**), the two consequent traps contributed at slightly higher degree, securing safer storage containment. This emphasizes that the geomechanics influence could not be excluded.

Strong discrepancy was however obvious in the syncline-shaped complex (**Fig. 10b**) due to a higher reservoir pressure system (**Fig. 4**), where trapping contributions in the homogeneous condition (black dotted lines in **Fig. 10b**) notably differ. In the initial period (<40 years), relative contribution from the structural and stratigraphic trap was higher when heterogeneity was included, implying an unavoidable heterogeneity nature when considers $CO_2$ storage containment. Similar results were also observed after such an initial period. Influence of geomechanics contributed to even less solubility trap development, which was due to lower reservoir pressure as anticipated. This also stresses that the geomechanics influence has to be taken into account when considering $CO_2$ storage containment via trapping mechanisms.

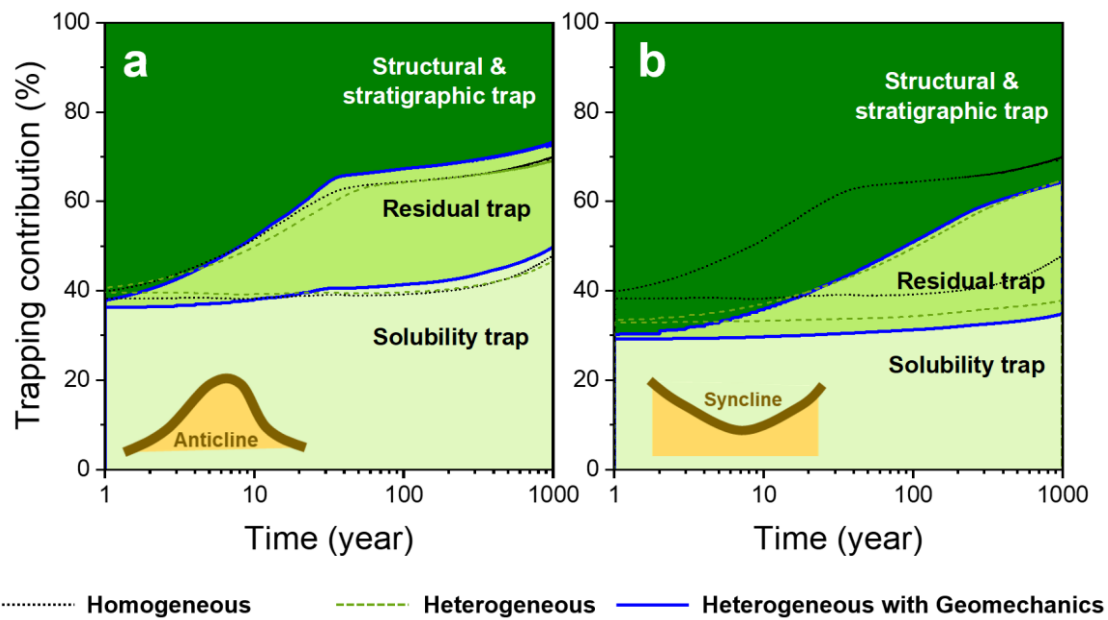

**Figure 10.** $CO_2$ trapping mechanism contributions over 1000 years since the injection started for both reservoir complexes simulated: anticline-shaped (a) and syncline-shaped (b). All simulated conditions consider geochemistry effect with further inclusion of: homogeneous effect (shown by black dotted lines); heterogeneous effect (shown by green dashed lines); and heterogeneous with geomechanics effect (shown by blue solid lines). Only the first three trapping mechanisms are reported, while the fourth of mineral trapping is negligible over the studied condition. Insets are illustrated graphics showing the reservoir shapes: anticline and syncline.

## 3.4 Sensitivity Analysis

As a result of the increase in reservoir pore pressure to be above critical threshold, the consequent decrease in effective stress could induce reservoir failure. Sensitivity analysis is therefore conducted to determine how responsive the reservoir pore pressure is on some input geomechanical parameters of the reservoir models. The considered parameters include reservoir porosity, permeability, Poisson's ratio, and Young's modulus.

The normalized sensitivity coefficient ($NSC$), used to assess the sensitivity of each parameter on the reservoir pore pressure, is defined as:[74]

$$NSC_i = \frac{\Delta Y}{\bar{Y}} \frac{\bar{X}_i}{\Delta X_i} \qquad (3)$$

where $\bar{Y}$ is the model output's nominal value at a nominal input model parameter $\bar{X}_i$. The change in the model output function caused by a change in $\Delta X_i$ in the input parameter $X_i$ is known as the variation $\Delta Y$. The four considered parameters are the model inputs ($X_i$), while the maximum reservoir pore pressure after the $CO_2$ injection is the output function $Y$. The numbers $X+$ and $X-$ are changes upon the base value of the input parameter $X$, considering $\pm 10\%$, while the values $Y+$ and $Y-$ are the according values of the output function $Y$. The higher the $NSC$, the more sensitive the input parameter.

Results of the sensitivity analysis for both anticline-shaped and syncline-shaped are reported in **Table 4** and **Table 5**, respectively. Very low values of ($<1 \times 10^{-2}$) $NSC$ were found with all considered input parameters in both complexes, implying negligible sensitivity of the parameters at the studied ranges to the output of reservoir pore pressure. This also suggests a good reservoir stability over the $CO_2$ injection design simulated.

**Table 4**. Results of the normalized sensitivity analysis ($NSC$) for anticline-shaped complex.

| Parameter | $X$ | $X+$ | $X-$ | $Y+$ | $Y-$ | $\Delta X_i$ | $\Delta Y_i$ | $NSC_i$ |
|---|---|---|---|---|---|---|---|---|
| Porosity (%) | 1.95 | 2.145 | 1.755 | 17156 | 17348 | 0.39 | 192 | $2.86 \times 10^{-3}$ |
| Permeability (mD) | 1.357 | 1.4927 | 1.2213 | 17317 | 17185 | 0.2714 | 132 | $1.97 \times 10^{-3}$ |
| Poisson's ratio | 0.25 | 0.275 | 0.225 | 17181 | 17150 | 0.05 | 31 | $4.62 \times 10^{-4}$ |
| Young's modulus (GPa) | 10.7 | 11.77 | 9.63 | 17203 | 17106 | 2.14 | 97 | $1.45 \times 10^{-3}$ |

**Table 5**. Results of the normalized sensitivity analysis ($NSC$) for syncline-shaped complex.

| Parameter | $X$ | $X+$ | $X-$ | $Y+$ | $Y-$ | $\Delta X_i$ | $\Delta Y_i$ | $NSC_i$ |
|---|---|---|---|---|---|---|---|---|
| Porosity (%) | 1.95 | 2.145 | 1.755 | 29474 | 30145 | 0.39 | 671 | $4.91 \times 10^{-3}$ |
| Permeability (mD) | 1.357 | 1.4927 | 1.2213 | 30828 | 29733 | 0.2714 | 1095 | $8.01 \times 10^{-3}$ |
| Poisson's ratio | 0.25 | 0.275 | 0.225 | 30018 | 29905 | 0.05 | 113 | $8.26 \times 10^{-4}$ |
| Young's modulus (GPa) | 10.7 | 11.77 | 9.63 | 30050 | 29849 | 2.14 | 201 | $1.47 \times 10^{-3}$ |

**4 Conclusion**

Numerical reservoir simulation for $CO_2$ storage in tight sandstone formation at Mae Moh Basin (Lampang, Thailand) has been performed, whereas geomechanics factor is heavily examined on its attribution to $CO_2$ plume migration and dynamic contribution of the storage trapping mechanisms, in addition to reservoir heterogeneity and geochemistry factors. Reservoir models of anticline-shaped and syncline-shaped at different depths were investigated, representing the actual basin geology. Main conclusions drawn from the simulated results are as follows:

- Strong reservoir pressure build-up was observed in the continuous injection period of 30 years with subsequent slow dissipation of the pressure to a stabilized level at the post-injection period. With injection rates designed, the pressure responses did not exceed reservoir fracture pressure for both complexes, demonstrating storage security. Heterogeneity induced slightly higher-pressure build-up due to limited fluid flow paths, but when incorporating geomechanics the reservoir pressures reduced because of coupling effects of pressure and pore space considered action upon.

- Geomechanics was found to limit $CO_2$ plume migration in both vertical and horizontal paths, while heterogeneity rather blocked the vertical direction and led to more lateral migration, as anticipated. Approaching a millennium, the plume gradually traveled upward after the injection but did not escape from the trap. Although both complexes showed similar plume migration behavior, those of syncline-shaped migrated to lesser extent owing to higher pressure as attributed to a deeper reservoir. Consideration of geomechanics also led to $CO_2$ plume traversing toward more constrained spaces and reduced leakage risks, highlighting a significant and rather neglected influence of geomechanical factor when considering $CO_2$ plume dynamics.

- $CO_2$ trapping mechanisms were observed and quantified while mineral trapping appeared to be negligible. Depending on the reservoir pressure as contributed to the reservoir depth, not the reservoir shape, geomechanics influenced the trapping mechanisms to different degrees. For a shallower reservoir, geomechanics contributed to slightly higher degrees of the residual and solubility trappings, implying a more secured storage containment after the injection. On the contrary, with deeper reservoirs, geomechanics substantially attributed to less solubility trapping, likely due to lower reservoir pressure.

With many criteria and constraints to be considered, geomechanics is one of the crucial factors that has to be assessed in order to implement CCS projects. Furthermore, the current study reveals a possibility of utilizing tight sandstone formations – not limited to specific areas, but any prospects globally – as storage reservoirs for $CO_2$ at gigaton scale, with potentially high levels storage containment and integrity.

**Declaration of competing interest**

The authors declare that they have no known competing financial interests or personal relationships that could have appeared to influence the work reported in this paper.

**Acknowledgements**

Financial support for this work is greatly acknowledged with contributions from: (i) the Electricity Generating Authority of Thailand [grant number 65-G201000-11-IO.SG03G3008639]; (ii) the National Science, Research and Innovation Fund (NSRF) via the Program Management Unit for Human Resources & Institutional Development, Research and Innovation [grant number B40G660032]; (iii) Frontier Research Grant from Faculty of Engineering, Chiang Mai University; and Chiang Mai University through Chiang Mai Research Center for Carbon Capture and Storage (Chiang Mai CCS) (No. RG 58/2566). Computer Modelling Group Ltd. (Canada) is gratefully acknowledged for providing the CMG simulator licenses.

**References**

(1)　UNFCCC. Matters Relating to the Global Stocktake under the Paris Agreement; UN Climate Change Conference, 2023.

(2)　IPCC. *IPCC Sixth Assessment Report - Synthesis Report*; 2023.

(3)　Yamada, K.; Ramon, B.; Fernandes, B.; Kalamkar, A.; Jeon, J.; Delshad, M.; Farajzadeh, R.; Sepehrnoori, K. Development of a Hydrate Risk Assessment Tool Based on Machine Learning for CO 2 Storage in Depleted Gas Reservoirs. *Fuel* **2024**, *357* (PA), 129670. https://doi.org/10.1016/j.fuel.2023.129670.

(4)　Krevor, S.; de Coninck, H.; Gasda, S. E.; Ghaleigh, N. S.; de Gooyert, V.; Hajibeygi, H.; Juanes, R.; Neufeld, J.; Roberts, J. J.; Swennenhuis, F. Subsurface Carbon Dioxide and Hydrogen Storage for a Sustainable Energy Future. *Nat. Rev. Earth Environ.* **2023**, *4* (2), 102–118. https://doi.org/10.1038/s43017-022-00376-8.

(5)　Krevor, S.; Blunt, M. J.; Benson, S. M.; Pentland, C. H.; Reynolds, C.; Al-Menhali, A.; Niu, B. Capillary Trapping for Geologic Carbon Dioxide Storage - From Pore Scale Physics to Field Scale Implications. *Int. J. Greenh. Gas Control* **2015**, *40*, 221–237. https://doi.org/10.1016/j.ijggc.2015.04.006.

(6)　Akai, T.; Saito, N.; Hiyama, M.; Okabe, H. Numerical Modelling on CO2 Storage Capacity in Depleted Gas Reservoirs. *Energies* **2021**, *14* (13). https://doi.org/10.3390/en14133978.

(7)　Temitope, A.; Sulemana, N.; Gomes, J. S.; Oppong, R. Statistical Uncertainty Analysis and Design Optimization of CO2 Trapping Mechanisms in Saline Aquifers. **2016**. https://doi.org/10.2523/iptc-18837-ms.

(8)　Ajayi, T.; Gomes, J. S.; Bera, A. A Review of CO2 Storage in Geological Formations Emphasizing Modeling, Monitoring and Capacity Estimation Approaches. *Pet. Sci.* **2019**, *16* (5), 1028–1063.

https://doi.org/10.1007/s12182-019-0340-8.

(9) Cooper, C. *A Technical Basis for Carbon Dioxide Storage*; 2009; Vol. 1. https://doi.org/10.1016/j.egypro.2009.01.226.

(10) Nghiem, L.; Shrivastava, V.; Kohse, B.; Hassam, M.; Yang, C. Simulation of Trapping Processes for CO2 Storage in Saline Aquifers. *Can. Int. Pet. Conf. 2009, CIPC 2009* **2009**, No. 8, 15–22. https://doi.org/10.2118/2009-156.

(11) Mackay, E. J. *Modelling the Injectivity, Migration and Trapping of CO2 in Carbon Capture and Storage (CCS)*; Woodhead Publishing Limited, 2013; Vol. 45. https://doi.org/10.1533/9780857097279.1.45.

(12) Zhan, J.; Soo, E.; Fogwill, A.; Cheng, S.; Cai, H.; Zhang, K.; Chen, Z. A Systematic Reservoir Simulation Study on Assessing the Feasibility of CO2 Sequestration in Liquid-Rich Shale Gas Reservoirs with Potential Enhanced Gas Recovery. *Offshore Technol. Conf. Asia 2018, OTCA 2018* **2018**.

(13) Petrusak, R.; Riestenberg, D.; Goad, P.; Schepers, K.; Pashin, J.; Esposito, R.; Trautz, R. World Class CO2 Sequestration Potential in Saline Formations, Oil and Gas Fields, Coal, and Shale: The US Southeast Regional Carbon Sequestration Partnership Has It All. *SPE Int. Conf. CO2 Capture, Storage, Util. 2009* **2009**, 136–153. https://doi.org/10.2118/126619-ms.

(14) AlRassas, A. M.; Ren, S.; Sun, R.; Thanh, H. V.; Guan, Z. CO2 Storage Capacity Estimation under Geological Uncertainty Using 3-D Geological Modeling of Unconventional Reservoir Rocks in Shahejie Formation, Block Nv32, China. *J. Pet. Explor. Prod.* **2021**, *11* (6), 2327–2345. https://doi.org/10.1007/s13202-021-01192-4.

(15) Ribeiro, A.; Sedaghat, M.; Honari, V.; Hurter, S. Impact of Injection Temperature on CO2 Storage in the Surat Basin, Eastern Australia. *Soc. Pet. Eng. - SPE Asia Pacific Oil Gas Conf. Exhib. 2020, APOG 2020* **2020**. https://doi.org/10.2118/202333-ms.

(16) Verma, Y.; Vishal, V.; Ranjith, P. G. Sensitivity Analysis of Geomechanical Constraints in CO2 Storage to Screen Potential Sites in Deep Saline Aquifers. *Front. Clim.* **2021**, *3* (October), 1–22. https://doi.org/10.3389/fclim.2021.720959.

(17) Li, N.; Feng, W.; Yu, J.; Chen, F.; Zhang, Q.; Zhu, S.; Hu, Y.; Li, Y. Recent Advances in Geological Storage: Trapping Mechanisms, Storage Sites, Projects, and Application of Machine Learning. *Energy and Fuels* **2023**. https://doi.org/10.1021/acs.energyfuels.3c01433.

(18) Voskov, D. V.; Henley, H.; Lucia, A. Fully Compositional Multi-Scale Reservoir Simulation of Various CO2 Sequestration Mechanisms. *Comput. Chem. Eng.* **2017**, *96*, 183–195. https://doi.org/10.1016/j.compchemeng.2016.09.021.

(19) Rashidi, M. R. A.; Dabbi, E. P.; Abu Bakar, Z. A.; Shahir Misnan, M.; Pedersen, C.; Wong, K. Y.; Sallehud-Din, T. M. M.; Azim Shamsudin, M.; Wo, S. A Field Case Study of Modelling the Environmental Fate of Leaked CO2 Gas in the Marine Environment for Carbon Capture and Storage CCS. *Soc. Pet. Eng. - SPE Asia Pacific Oil Gas Conf. Exhib. 2020, APOG 2020* **2020**. https://doi.org/10.2118/202394-ms.

(20) Mohamed, I. M.; Nasr-El-Din, H. A. Formation Damage Due to CO2 Sequestration in Deep Saline Carbonate Aquifers. *Proc. - SPE Int. Symp. Form. Damage Control* **2012**, *1* (Gruber 1996), 319–350.

https://doi.org/10.2118/151142-ms.

(21) Jackson, S. J.; Krevor, S. Small-Scale Capillary Heterogeneity Linked to Rapid Plume Migration During CO2 Storage. **2020**.

(22) Harris, C.; Jackson, S. J.; Benham, G. P.; Krevor, S.; Muggeridge, A. H. The Impact of Heterogeneity on the Capillary Trapping of CO2 in the Captain Sandstone. *Int. J. Greenh. Gas Control* **2021**, *112*, 103511. https://doi.org/10.1016/j.ijggc.2021.103511.

(23) Alsayah, A.; Rigby, S. P. Coupled Multiphase Flow, Geochemical, and Geomechanical Modelling of the Impact of Shale Interlayers on CO2 Migration. *Geoenergy Sci. Eng.* **2023**, *229* (June), 212101. https://doi.org/10.1016/j.geoen.2023.212101.

(24) Fang, X.; Lv, Y.; Yuan, C.; Zhu, X.; Guo, J.; Liu, W.; Li, H. Effects of Reservoir Heterogeneity on CO2 Dissolution Efficiency in Randomly Multilayered Formations. *Energies* **2023**, *16* (13). https://doi.org/10.3390/en16135219.

(25) Al-Khdheeawi, E. A.; Vialle, S.; Barifcani, A.; Sarmadivaleh, M.; Iglauer, S. Impact of Reservoir Wettability and Heterogeneity on CO2-Plume Migration and Trapping Capacity. *Int. J. Greenh. Gas Control* **2017**, *58*, 142–158. https://doi.org/10.1016/j.ijggc.2017.01.012.

(26) Rasheed, Z.; Raza, A.; Gholami, R.; Rabiei, M.; Ismail, A.; Rasouli, V. A Numerical Study to Assess the Effect of Heterogeneity on CO2 Storage Potential of Saline Aquifers. *Energy Geosci.* **2020**, *1* (1–2), 20–27. https://doi.org/10.1016/j.engeos.2020.03.002.

(27) Gershenzon, N. I.; Soltanian, M.; Ritzi, R. W.; Dominic, D. F. Influence of Small Scale Heterogeneity on CO2 Trapping Processes in Deep Saline Aquifers. *Energy Procedia* **2014**, *59*, 166–173. https://doi.org/10.1016/j.egypro.2014.10.363.

(28) Jackson, S. J.; Krevor, S. Small-Scale Capillary Heterogeneity Linked to Rapid Plume Migration During.Pdf. *Geophys. Res. Lett.* **2020**.

(29) Al-Khdheeawi, E. A.; Vialle, S.; Barifcani, A.; Sarmadivaleh, M.; Iglauer, S. Effect of Wettability Heterogeneity and Reservoir Temperature on CO2 Storage Efficiency in Deep Saline Aquifers. *Int. J. Greenh. Gas Control* **2018**, *68* (November 2017), 216–229. https://doi.org/10.1016/j.ijggc.2017.11.016.

(30) Chidambaram, P.; Tiwari, P. K.; Patil, P. A.; Mohd Ali, S. S.; Amin, S. M.; Tewari, R. D.; Tan, C. P.; Widyanita, A.; Hamid, K. M. B. A. Importance of 3-Way Coupled Modelling for Carbon Dioxide Sequestration in Depleted Reservoir. *Proc. - SPE Annu. Tech. Conf. Exhib.* **2021**, *2021-Septe*. https://doi.org/10.2118/206156-MS.

(31) Khan, S.; Khulief, Y.; Al-Shuhail, A.; Bashmal, S.; Iqbal, N. The Geomechanical and Fault Activation Modeling during Co2 Injection into Deep Minjur Reservoir, Eastern Saudi Arabia. *Sustain.* **2020**, *12* (23), 1–17. https://doi.org/10.3390/su12239800.

(32) Jun, S.; Song, Y.; Wang, J.; Weijermars, R. Formation Uplift Analysis during Geological CO2-Storage Using the Gaussian Pressure Transient Method: Krechba (Algeria) Validation and South Korean Case Studies. *Geoenergy Sci. Eng.* **2023**, *221* (December 2022), 211404. https://doi.org/10.1016/j.geoen.2022.211404.

(33) Chiaramonte, L.; White, J. a; Trainor-guitton, W. Probabilistic Geomechanical Analysis of Compartmentalization at the Snøhvit CO2 Sequestration Project Laura. *AGU J. Geophys. Res. Solid*

*Earth* **2014**, *120*, 1195–1209. https://doi.org/10.1002/2014JB011376.Received.

(34) Vilarrasa, V.; Makhnenko, R.; Gheibi, S. Geomechanical Analysis of the Influence of CO2 Injection Location on Fault Stability. *J. Rock Mech. Geotech. Eng.* **2016**, *8* (6), 805–818. https://doi.org/10.1016/j.jrmge.2016.06.006.

(35) Song, Y.; Jun, S.; Na, Y.; Kim, K.; Jang, Y.; Wang, J. Geomechanical Challenges during Geological CO2 Storage: A Review. *Chem. Eng. J.* **2023**, *456* (December 2022), 140968. https://doi.org/10.1016/j.cej.2022.140968.

(36) Khan, S.; Khulief, Y. A.; Al-Shuhail, A. Mitigating Climate Change via CO2 Sequestration into Biyadh Reservoir: Geomechanical Modeling and Caprock Integrity. *Mitig. Adapt. Strateg. Glob. Chang.* **2019**, *24* (1), 23–52. https://doi.org/10.1007/s11027-018-9792-1.

(37) Gholami, R.; Raza, A. CO2 Sequestration in Sandstone Reservoirs: How Does Reactive Flow Alter Trapping Mechanisms? *Fuel* **2022**, *324* (PC), 124781. https://doi.org/10.1016/j.fuel.2022.124781.

(38) Poland, N. Numerical Simulations of Carbon Dioxide Storage in Selected Geological Structures In. **2022**, *10* (November 2021), 1–19. https://doi.org/10.3389/fenrg.2022.827794.

(39) Chu, A. K.; Benson, S. M.; Wen, G. Deep-Learning-Based Flow Prediction for CO 2 Storage in Shale – Sandstone Formations. **2023**.

(40) Thanasaksukthawee, V.; Santha, N.; Saenton, S.; Tippayawong, N.; Jaroonpattanapong, P.; Foroozesh, J.; Tangparitkul, S. Relative CO2Column Height for CO2Geological Storage: A Non-Negligible Contribution from Reservoir Rock Characteristics. *Energy and Fuels* **2022**, *36* (7), 3727–3736. https://doi.org/10.1021/acs.energyfuels.1c04398.

(41) EGAT. *EGAT Sustainability Report*; 2022.

(42) Somprasong, K.; Hutayanon, T.; Jaroonpattanapong, P. Using Carbon Sequestration as a Remote-Monitoring Approach for Reclamation's Effectiveness in the Open Pit Coal Mine: A Case Study of Mae Moh, Thailand. *Energies* **2024**, *17* (1). https://doi.org/10.3390/en17010231.

(43) Chaodumrong, P. Stratigraphy, Sedimentology and Tectonic Setting of the Lampang Group, Central North Thailand, University of Tasmania, 1992.

(44) Netsakkasame, S.; Boontun, A.; Saenton, S. Groundwater Flow Modeling for Pit Wall Stability and Floor Heave Analyses: A Case Study of Mae Moh Mine. *Tjyybjb.Ac.Cn* **2021**, *27* (2), 58–66.

(45) Chaodumrong, P.; Burret, C. F. *Stratigraphy of the Lampang Group in Central North Thailand: New Version*; 1997.

(46) Computer Modelling Group Ltd. *GEM User's Guide - Advanced Compositional and GHG Reservoir Simulator*; 2009.

(47) Computer Modelling Group Ltd. *CMG GEM Manual*; 2023.

(48) Kamali, F.; Hussain, F.; Cinar, Y.; South, N. Study of Co-Optimizing CO 2 Storage and CO 2 Enhanced Oil Recovery. **2015**, No. December, 1–11.

(49) Venkatraman, A.; Dindoruk, B.; Elshahawi, H.; Lake, L. W.; Johns, R. T. Modeling Effect of Geochemical Reactions on Real-Reservoir-Fluid Mixture during Carbon Dioxide Enhanced Oil Recovery. *SPE J.* **2017**, *22* (5), 1519–1529. https://doi.org/10.2118/175030-pa.

(50) Ramadhan, R.; Novriansyah, A.; Erfando, T.; Tangparitkul, S.; Daniati, A.; Permadi, A. K.; Abdurrahman, M. Heterogeneity Effect on Polymer Injection: A Study of Sumatra Light Oil. *Sci. Contrib. Oil Gas* **2023**, *46* (1), 39–52. https://doi.org/10.29017/SCOG.46.1.1322.

(51) Akai, T.; Kuriyama, T.; Kato, S.; Okabe, H. Numerical Modelling of Long-Term CO2 Storage Mechanisms in Saline Aquifers Using the Sleipner Benchmark Dataset. *Int. J. Greenh. Gas Control* **2021**, *110*. https://doi.org/10.1016/j.ijggc.2021.103405.

(52) Johnson, Kaj, M. Growth of Fault-Cored Anticlines by Flexural Slip Folding Analysis by Boundary Element. *J. Geophys. Res. Solid Earth* **2018**. https://doi.org/10.1002/2017JB014867 Key.

(53) Ramadhan, R.; Abdurrahman, M.; Bissen, R.; Maneeintr, K. Numerical Simulation of Potential Site for CO2 Sequestration in a Depleted Oil Reservoir in Northern Thailand. *Energy Reports* **2023**, *9* (S11), 524–528. https://doi.org/10.1016/j.egyr.2023.09.096.

(54) Maneeintr, K.; Ruanman, N.; Juntarasakul, O. Assessment of CO2 Geological Storage Potential in a Depleted Oil Field in the North of Thailand. *Energy Procedia* **2017**, *141*, 175–179. https://doi.org/10.1016/j.egypro.2017.11.033.

(55) Zhang, J. J. In Situ Stress Estimate. In *Applied Petroleum Geomechanics*; 2019; pp 187–232. https://doi.org/10.1016/b978-0-12-814814-3.00006-x.

(56) Tingay, M. R. P.; Morley, C. K.; Hillis, R. R.; Meyer, J. Present-Day Stress Orientation in Thailand's Basins. *J. Struct. Geol.* **2010**, *32* (2), 235–248. https://doi.org/10.1016/j.jsg.2009.11.008.

(57) Alrassas, A. M.; Vo Thanh, H.; Ren, S.; Sun, R.; Al-Areeq, N. M.; Kolawole, O.; Hakimi, M. H. CO2Sequestration and Enhanced Oil Recovery via the Water Alternating Gas Scheme in a Mixed Transgressive Sandstone-Carbonate Reservoir: Case Study of a Large Middle East Oilfield. *Energy and Fuels* **2022**. https://doi.org/10.1021/acs.energyfuels.2c02185.

(58) Hubbert, M, K.; Willis, D. G. Mechanics of Hydraulic Fracturing. *AIME Pet. Trans.* **1957**, *210*, 153–168. https://doi.org/10.1016/B978-0-12-822195-2.00020-6.

(59) Oruganti, Y. D.; Bryant, S. L. Pressure Build-up during CO2 Storage in Partially Confined Aquifers. *Energy Procedia* **2009**, *1* (1), 3315–3322. https://doi.org/10.1016/j.egypro.2009.02.118.

(60) Terzaghi, K. *Theoretical Soil Mechanics*; 1943; Vol. 13. https://doi.org/10.1680/geot.1963.13.4.267.

(61) Li, S.; Wang, P.; Wang, Z.; Cheng, H.; Zhang, K. Strategy to Enhance Geological CO2 Storage Capacity in Saline Aquifer. *Geophys. Res. Lett.* **2023**, *50* (3). https://doi.org/10.1029/2022GL101431.

(62) Haghighat, S. A.; Mohaghegh, S. D.; Gholami, V.; Shahkarami, A.; Moreno, D. Using Big Data and Smart Field Technology for Detecting Leakage in a CO2 Storage Project. *Proc. - SPE Annu. Tech. Conf. Exhib.* **2013**, *1*, 815–821. https://doi.org/10.2118/166137-ms.

(63) Alhotan, M.; Jia, C.; Alsousy, A.; Delshad, M.; Sepehrnoori, K. Impact of Permeability Heterogeneity Coupled with Well Placement Strategy on Underground Hydrogen Storage Reservoir Simulation. *SPE Middle East Oil Gas Show Conf. MEOS, Proc.* **2023**. https://doi.org/10.2118/213257-MS.

(64) Kim, K.; Vilarrasa, V.; Makhnenko, R. Y. CO2 Injection Effect on Geomechanical and Flow Properties of Calcite-Rich Reservoirs. *Fluids* **2018**, *3* (3). https://doi.org/10.3390/fluids3030066.

(65) Agada, S.; Jackson, S.; Kolster, C.; Dowell, N. Mac; Williams, G.; Vosper, H.; Williams, J.; Krevor, S. The Impact of Energy Systems Demands on Pressure Limited CO2 Storage in the Bunter Sandstone of the UK Southern North Sea. *Int. J. Greenh. Gas Control* **2017**, *65* (September), 128–136. https://doi.org/10.1016/j.ijggc.2017.08.014.

(66) Khudaida, K. J.; Das, D. B. A Numerical Analysis of the Effects of Supercritical CO2 Injection on CO2 Storage Capacities of Geological Formations. *Clean Technol.* **2020**, *2* (3), 333–364. https://doi.org/10.3390/cleantechnol2030021.

(67) Sohal, M. A.; Le Gallo, Y.; Audigane, P.; de Dios, J. C.; Rigby, S. P. Effect of Geological Heterogeneities on Reservoir Storage Capacity and Migration of CO2 Plume in a Deep Saline Fractured Carbonate Aquifer. *Int. J. Greenh. Gas Control* **2021**, *108* (March). https://doi.org/10.1016/j.ijggc.2021.103306.

(68) Singh, M.; Mahmoodpour, S.; Schmidt-hattenberger, C.; Sass, I.; Drews, M. Influence of Reservoir Heterogeneity on Simultaneous Geothermal Energy Extraction and CO 2 Storage. **2024**.

(69) Hansen, S. K.; Tao, Y.; Karra, S. Impacts of Permeability Heterogeneity and Background Flow on Supercritical CO2 Dissolution in the Deep Subsurface. *Water Resour. Res.* **2023**, *59* (11). https://doi.org/10.1029/2023WR035394.

(70) Yong, W. P.; Azahree, A. I.; Ali, S. S. M.; Azuddin, F. J.; Amin, S. M. A New Modelling Approach to Simulate CO2 Movement and Containment Coupled with Geochemical Reactions and Geomechanical Effects for an Offshore CO2 Storage in Malaysia. *Soc. Pet. Eng. - SPE Eur. Featur. 81st EAGE Conf. Exhib. 2019* **2019**. https://doi.org/10.2118/195432-ms.

(71) Nordbotten, J. A. N. M.; Celia, M. A.; Bachu, S. Injection and Storage of CO 2 in Deep Saline Aquifers : Analytical Solution for CO 2 Plume Evolution During Injection. *Transp. Porous Media* **2005**, 339–360. https://doi.org/10.1007/s11242-004-0670-9.

(72) Nordbotten, J. A. N. M.; Celia, M. A. Similarity Solutions for Fluid Injection into Confined Aquifers. *J Fluid Mech* **2006**, *561*, 307–327. https://doi.org/10.1017/S0022112006000802.

(73) Metz, B.; Davidson, O.; De Coninck, H.; Loos, M.; Meyer, L. *IPCC Special Report on Carbon Dioxide Capture and Storage*; 2005; Vol. 58. https://doi.org/10.1016/bs.ache.2021.10.005.

(74) Masi, M.; Fogliani, S.; Carrà, S. Sensitivity Analysis on Indium Phosphide Liquid Encapsulated Czochralski Growth. *Cryst. Res. Technol.* **1999**, *34* (9), 1157–1167. https://doi.org/10.1002/(SICI)1521-4079(199911)34:9<1157::AID-CRAT1157>3.0.CO;2-V.